\documentclass[english,twocolumn,superscriptaddress,prl,aps,showpacs]{revtex4}
\usepackage[T1]{fontenc}
\usepackage[latin1]{inputenc}
\usepackage{graphicx}
\usepackage{amssymb}

\makeatletter

\providecommand{\boldsymbol}[1]{\mbox{\boldmath $#1$}}

\usepackage{bm}

\usepackage{dcolumn}

\usepackage{bm}

\usepackage{endnotes}

\usepackage{amsfonts}

\providecommand{\U}[1]{\protect\rule{.1in}{.1in}}
\makeatletter
\makeatletter
\makeatother
\makeatother

\usepackage{babel}
\makeatother

\begin{document}

\author{Alexey A. Kovalev}

\affiliation{Department of Physics, Texas A\&M University, College Station, TX
77843-4242, USA}

\author{Liviu P. Z\^arbo }

\affiliation{Department of Physics, Texas A\&M University, College Station, TX
77843-4242, USA}

\author{Y. Tserkovnyak}

\affiliation{Department of Physics and Astronomy, University of California, Los
Angeles, California 90095, USA}

\author{G. E. W. Bauer}

\affiliation{Kavli Institute of NanoScience, Delft University of Technology, Lorentzweg
1, 2628CJ Delft, The Netherlands}

\author{Jairo Sinova}

\affiliation{Department of Physics, Texas A\&M University, College Station, TX
77843-4242, USA}

\title{Nanomechanical Spin-Polarizer }

\begin{abstract}
Torsional oscillations of a free-standing semiconductor beam are shown
to cause spin-dependent oscillating potentials that spin-polarize
an applied charge current in the presence of intentional or disorder
scattering potentials. We propose several realizations of mechanical
spin generators and manipulators based on this piezo-spintronic effect.
\end{abstract}

\date{\today{}}


\maketitle
The field of spintronics comprises the search for novel logic and
sensing devices that employ the electron spin degree of freedom by
(excess) spin generation and manipulation \citep{Zutic:apr2004}.
In the conventional approach spins are injected into normal conductors
by ferromagnetic metals using an applied electrical bias. An alternative
method is the spin-pumping by a moving magnetization \citep{Brataas:aug2002,Tserkovnyak:dec2005}.
A net-spin generation is possible without involving ferromagnets at
all by making use of the spin-orbit interaction, e.g. by a time-dependent
gate acting on a two-dimensional electron gas \citep{Governale:oct2003,Tang:may2005}, or the
spin Hall effect \citep{SHE}.

Spin-transfer by spin-flip scattering in metal structures causes mechanical
torques \citep{Fulde:1998,Mohanty:nov2004,Kovalev:jan2007}. Mal'shukov
\textit{et. al.} \citep{Mal'shukov:sep2005} predict that a spin-polarized
current can induce torsional vibrations in a semiconductor beam by
strain-induced spin-orbit interaction. The same authors speculate
about a possible reverse effect, \textit{viz.} that mechanical motion
could induce a spin-polarized current.

In this Letter, we propose a nano-electro-mechanical system (NEMS)
that generates spins by the coupling to torsional oscillations of
a free-standing semiconductor bridge/beam/rod that is actuated, e.g.,
by magnetomotive \citep{Mohanty:nov2004,Huang:nov2005}, electrostatic
\citep{Fennimore:jul2003} or piezoelectric \citep{Masmanidis:aug2007}
forces. Subject to an oscillating strain, the spin-orbit interaction
in the semiconductor generates a spin splitting which, in the presence
of a bias, leads to a spin current. In analogy with piezo-electricity,
in which elastic strain induces free charges, this can be termed a
\emph{piezo-spintronic} effect. We illustrate the physical principle
by a conducting wire in the electric quantum limit in which only a
single quantized subband is occupied. Subsequently, we generalize
the results to the multi-channel case. We also demonstrate by numerical
simulations that the effect survives the disorder that can be expected
in real systems and discuss the conditions under which it can be observed.

\begin{figure}[t]
\centerline{\includegraphics[scale=0.3]{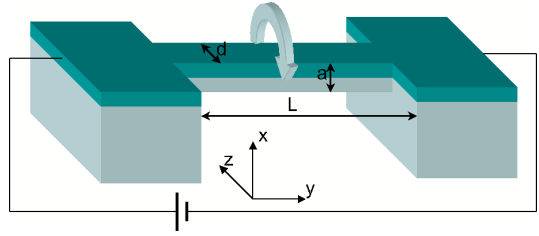}}
\caption{A beam consisting of a semiconductor and insulator parts (semiconductor
layer of thickness $a/2$ is on top of insulator layer of the same
thickness) is excited by some external source into torsional oscillations.
A voltage is applied over the device in order to detect the mechanically
induced spin splitting.}
\label{RodDraw}
\end{figure}

Let us consider a beam with length $L$ and rectangular cross section
of width $d$ and thickness $a$ ($L\gg d(a)$) (see Fig \ref{RodDraw})
that connects two semi-infinite conducting reservoirs. Results can
be easily generalized to axially symmetric rods such as catalytically
grown nanowires \citep{Duan:jan2001}. 
The conducting material is
a semiconductor that is grown on top of a dielectric. Without loss
of generality we assume here a structure consisting of a conducting
medium on top of an insulator both being $a/2$ thick. %
The leading modification of the conduction band Hamiltonian of a semiconductor
due to lattice strain reads \citep{Pikus:1984}: \begin{equation}
\begin{array}{c}
\widehat{H}_{SO}={\displaystyle \frac{\hbar^{2}}{2m^{\ast}}}\left\{ \gamma\left[\sigma_{y}(u_{xy}k_{x}-u_{yz}k_{z})+\sigma_{z}(u_{yz}k_{y}\right.\right.\\
\\\left.-u_{zx}k_{x})+\sigma_{x}(u_{zx}k_{z}-u_{xy}k_{y})\right]+\beta\left[\sigma_{y}k_{y}(u_{zz}-u_{xx})\right.\\
\\\left.+\left.\sigma_{z}k_{z}(u_{xx}-u_{yy})+\sigma_{x}k_{x}(u_{yy}-u_{zz})\right]+H.c.\right\}, \end{array}\label{ham}\end{equation}
 where $m^{\ast}$ is the effective mass, $u_{ij}$ are elements of
the strain tensor, and $\sigma_{i}$ are Pauli matrices.
By focusing on narrow-gap semiconductors, we may disregard the terms
proportional to the small parameter $\beta$ \citep{Pikus:1984}.

We are interested in the lowest energy vibrations of the beam that
can be described by an isotropic elastic continuum model \citep{Landau:1986}.
Elastic flexural (bending) modes cause only diagonal $u_{ii}$ strains
that couple to the electrons only via the small $\beta$ term (see
Eq. (\ref{ham})) \citep{Mal'shukov:sep2005}. The strain due to torsional
(twisting) vibrations is given by \citep{Landau:1986}:\begin{equation}
u_{zy}=\tau(y,t)\frac{\partial\chi}{\partial x};\; u_{xy}=-\tau(y,t)\frac{\partial\chi}{\partial z};\; u_{zx}=0,\end{equation}
 where $\tau(y,t)=\partial\varphi/\partial y$ is the derivative of
the torsion angle $\varphi$ with respect to $y$. The function $\chi$
characterizes the cross-section geometry of the beam and depends here
only on $x$ and $z$. It satisfies the equation $\Delta\chi=-1$
with vanishing boundary conditions \citep{Landau:1986}. We adopt
the thin plate geometry $a\ll d$ which leads to $\chi(x,z)\approx-(x^{2}-a^{2}/4)/2$
and the Hamiltonian:\begin{equation}
\widehat{H}_{SO}={\displaystyle \frac{\hbar^{2}}{2m^{\ast}}}\left[\gamma x\tau(y,t)\left(\sigma_{y}k_{z}-\sigma_{z}k_{y}\right)+H.c.\right].\label{ham1}\end{equation}
 Eq. (\ref{ham1}) is similar to a Rashba spin-orbit Hamiltonian;
however, electrons can move here in three dimensions and the coupling
strength is time and position dependent.

We now turn to the lowest electronic subband limit, disregarding intrinsic
spin-orbit interaction, e.g. Rashba type, 
and assuming that the strain induced perturbation is
weak. The free-electron lowest energy states read $\Psi(x,y,z)=R_{0}(x,z)\Phi(y),$
where $R_{0}(x,z)\sim\sin(\pi z/d)\sin(2\pi x/a)$ is the lowest subband
and $\Phi(y)$ is a spinor function. The projected one-dimensional
Hamiltonian then reads: \begin{equation}
\widehat{H}_{1D}(y)={\displaystyle \frac{\hbar^{2}}{2m^{\ast}}\left(k_{y}-{\displaystyle \frac{\gamma\tau(y,t)a}{4}}\sigma_{z}\right)^{2}}+V(y),\label{1DHam}\end{equation}
 where $V(y)$ is the potential due to impurities and we disregarded
terms $\backsim(\gamma\tau)^{2}$. Electrons with up and down spins
turn out to be uncoupled and subject to effective vector potentials
of opposite sign, $\mathbf{A}=\pm\hbar{\displaystyle \frac{\gamma\tau(y,t)a}{4}}\mathbf{y}$.
Since $\boldsymbol{\nabla}\times\mathbf{A}=0$, this vector potential
does not describe an effective magnetic but a spin-dependent \emph{electric}
field:\begin{equation}
\mathbf{E}=-\sigma_{z}{\displaystyle \frac{\hbar\gamma a}{4}}\frac{\partial\tau}{\partial t}\mathbf{y}.\label{electric}\end{equation}
The equation of motion for the torsional angle $\varphi\left(y,t\right)$
of the beam reads\begin{equation}
C\frac{\partial^{2}\varphi}{\partial y^{2}}-\rho I\frac{\partial^{2}\varphi}{\partial t^{2}}=0,\label{Mdynamics}\end{equation}
 where $I=\int(z^{2}+x^{2})dzdx\simeq ad^{3}/12$ is the moment of
inertia of the cross-section about its center of mass, $\rho$ the
mass density and $C$ is an elastic constant defined by the shape
and material of the cantilever. $C=\frac{1}{3}\mu da^{3}$ for a plate
with 
$a\ll d$, and $\mu$
is the Lamé constant. The general solution of Eq. (\ref{Mdynamics})
is a plane wave $\varphi=\varphi_{0}e^{i\omega t\pm iky}$, where
$k=\omega/c$ is the wave number, $c=2c_{\mathrm{t}}a/d=\sqrt{C/(\rho I)}$,
and $c_{\mathrm{t}}=\sqrt{\mu/\rho}$ is the sound velocity.
Throughout this paper, we consider a doubly-clamped beam in which
the lowest harmonic $\varphi=\varphi_{0}\sin(ky)\sin(\omega t)$ is
excited, where $\omega=ck$ and $k=\pi/L$ is the wave number (see
Fig. \ref{RodDraw}). The standing mechanical wave creates an oscillating
electric field $E=\partial\mathbf{A}/\partial t$ that is exactly
out of phase for spin-up and spin-down electrons. In the Born-Oppenheimer
approximation, the strain induces a parametric potential $U(y)={\displaystyle \frac{\hbar\gamma a\omega\varphi_{0}}{4}}\sin({\displaystyle \frac{\pi}{L}}y)\cos(\omega t)$
(cf. Eq. (\ref{electric})) that is adiabatically followed by the
electrons.

\begin{figure}[t]
\centerline{\includegraphics[width=0.8\columnwidth]{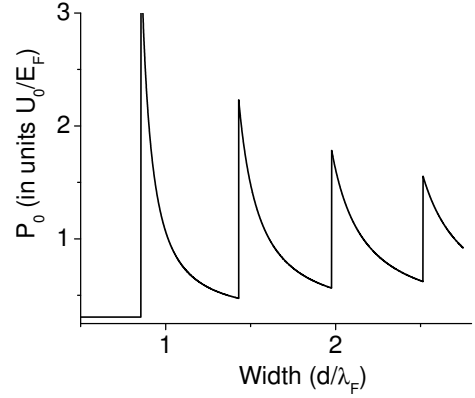}}
\vskip -0.5 cm
\caption{Maximum spin-polarization in units of the effective splitting $U_{0}/E_{F}$
as a function of the width of the electron waveguide in Fig. \ref{RodDraw}.
The frequency of the mechanical oscillations $\omega=10\mbox{GHz}$
which corresponds to $U_{0}/E_{F}=2\times10^{-4}$ (length of the
rod $L=1\mu\mbox{m}$, Fermi length $\lambda_{F}=30\mbox{nm}$, the
delta function strength $\upsilon^{2}m^{*}/\hbar^{2}=0.4E_{F}$ and
$m^{*}=0.06m$). }
\label{APolarization}
\vskip -0.5 cm

\end{figure}
The strain-induced potentials $\pm U(y)$ vary only slowly and do
not yet spin-polarize a charge current significantly. However, defect
scattering strongly amplifies the piezo-spintronic effect, as illustrated
now by a single short-range potential scatterer $V(y)=\upsilon\delta(y-L/2)$
located in the middle of the beam. Disregarding the small intrinsic
effect caused by $\pm U(y),$\ the probability that an electron with
Fermi wave number $k_{F}$ is transmitted by the scatterer reads:\begin{equation}
T_{\uparrow(\downarrow)}=\frac{2(\hbar^{2}k_{F}^{2}/2m^{\ast}\pm U_{0}\cos(\omega t))}{2(\hbar^{2}k_{F}^{2}/2m^{\ast}\pm U_{0}\cos(\omega t))+{\displaystyle \upsilon^{2}m^{\ast}/\hbar^{2}}},\end{equation}
 where $U_{0}={\displaystyle \frac{\hbar\gamma a\omega\varphi_{0}}{4}}$.
According to the Landauer conductance formula, the spin polarization
of a charge current becomes:\begin{equation}
P=\frac{T_{\uparrow}-T_{\downarrow}}{T_{\uparrow}+T_{\downarrow}}=\frac{U_{0}(\upsilon^{2}m^{\ast}/\hbar^{2})\cos(\omega t)}{E_{F}(2E_{F}+\upsilon^{2}m^{\ast}/\hbar^{2})-2U_{0}^{2}\cos^{2}(\omega t)},\label{Polarization}\end{equation}
 where the Fermi energy $E_{F}=\hbar^{2}k_{F}^{2}/2m^{\ast}$. The
spin-polarization oscillates in time with the beam vibration frequency.

Generalization to a multichannel wire is facilitated by the following
gauge transformation \begin{equation}
\psi=e^{if(x,y,t){\displaystyle \widehat{\sigma}_{z}}}\psi^{\prime}\label{Guage}\end{equation}
 with $e^{if(x,y,t){\displaystyle \widehat{\sigma}_{z}}}={\displaystyle \widehat{1}}\cos(f)+i{\displaystyle \widehat{\sigma}_{z}}\sin(f)$,
which leads to the transformed Hamiltonian:\begin{equation}
\begin{array}{c}
\widehat{H}_{SO}^{\prime}=e^{-if(x,y,t){\displaystyle \widehat{\sigma}_{z}}}\widehat{H}_{SO}e^{if(x,y,t){\displaystyle \widehat{\sigma}_{z}}}+\hbar{\displaystyle \frac{\partial f(x,y,t)}{\partial t}}\widehat{\sigma}_{z}\end{array},\label{TransfHam}\end{equation}
\begin{equation}
\begin{array}{c}
\widehat{H}={\displaystyle \frac{\hbar^{2}}{2m^{\ast}}\left(k_{x}+{\displaystyle \frac{\partial f(x,y,t)}{\partial x}}\widehat{\sigma}_{z}\right)^{2}}\\
+{\displaystyle \frac{\hbar^{2}}{2m^{\ast}}\left(k_{y}-\gamma\tau(y,t)x\sigma_{z}+{\displaystyle \frac{\partial f(x,y,t)}{\partial y}}\widehat{\sigma}_{z}\right)^{2}}\\
+{\displaystyle \frac{\hbar^{2}}{2m^{\ast}}\left(k_{z}+\gamma\tau(y,t)x(-\sin(2f)\sigma_{y}+\cos(2f)\sigma_{x})\right)^{2}}\\
+\hbar{\displaystyle \frac{\partial f(x,y,t)}{\partial t}}\widehat{\sigma}_{z}\end{array}\end{equation}
 It is convenient to choose ${\displaystyle \frac{\partial f(x,y,t)}{\partial y}}=\gamma\tau(y,t)a/4$.
We allow many occupied subbands along the $z$ axis but restrict considerations
to the lowest subband along the $x$ axis, which is the case for a
laterally weakly confined two-dimensional electron gas. After projecting
Eq. (\ref{TransfHam}) to the lowest mode in the $x$ direction, we
obtain the following two-dimensional Hamiltonian: \begin{equation}
\begin{array}{c}
\widehat{H}_{2D}={\displaystyle \frac{\hbar^{2}k_{y}^{2}}{2m^{\ast}}}+{\displaystyle \frac{\hbar^{2}}{2m^{\ast}}\biggl[k_{z}+{\displaystyle \frac{\gamma\tau(y,t)a}{4}}\Bigl(-\sin(2f)\sigma_{y}}\\
+\cos(2f)\sigma_{x}\Bigr)\biggr]^{2}+U(y)\widehat{\sigma}_{z}+V(y,z)\end{array},\label{Rashba}\end{equation}
 where $U(y)=\hbar{\displaystyle \frac{\partial f(y,t)}{\partial t}}={\displaystyle \frac{\hbar\omega L}{l_{so}}}\sin({\displaystyle \frac{\pi}{L}}y)\cos(\omega t)$,
${\displaystyle \frac{\gamma\tau(y,t)a}{4}}={\displaystyle \frac{\pi}{l_{so}}}\cos({\displaystyle \frac{y\pi}{L}})\sin(\omega t)$.
$l_{so}={\displaystyle \frac{4L}{\gamma\varphi_{0}a}}$ can be interpreted
as a spin precession length and $V(y,z)$ describes two-dimensional
disorder scattering. $\tau(y,t)$ is here still arbitrary, but we
limit our attention to the lowest vibrational mode as before.

\begin{figure}[t]
\centerline{\includegraphics[width=0.9\columnwidth]{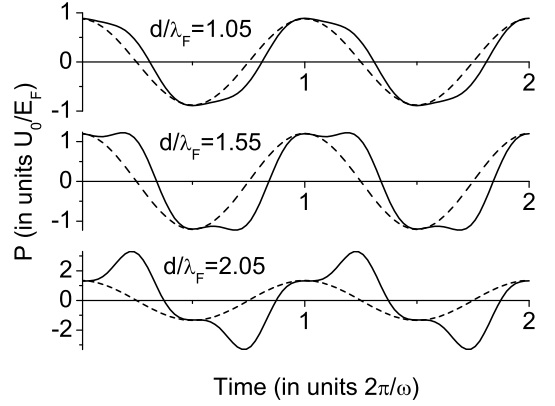}}
\vskip -0.5 cm
\caption{Spin-current polarization in units of the splitting parameter $U_{0}/E_{F}$
as a function of time for the torsional oscillations of the rod in
Fig. \ref{RodDraw}. The dashed line gives the results according to
Eq. (\ref{Polarization1}), whereas the bold line represent the results
of numerical simulations based on the tight binding model. Parameters
are the same as in Fig. \ref{APolarization}. }
\vskip -0.5 cm
\label{TPolarization}
\end{figure}

The terms proportional to $\sigma_{x(y)}k_{z}$ in Eq. (\ref{Rashba})
induce subband transitions; however, these do not affect transport
when the precession length $l_{so}$ is larger than the width of the
channel. In the limit of a long and narrow beam, we may again treat
the time dependence of the Hamiltonian Eq. (\ref{Rashba}) parametrically
in terms of the frequency $\omega$. In the limit $\bigtriangleup_{so} {\displaystyle \frac{d}{l_{so}}}\ll U_{0}={\displaystyle \frac{\hbar \omega L}{l_{so}}}$,
we can further simplify Eq. (\ref{Rashba}) by disregarding subband
transitions. A simple short-range wall potential $V(y,z)=\upsilon\delta(y-L/2)$
does not lead to subband transitions either. Our system then reduces
to a collection of independent channels, which lead to a total spin-current
polarization\begin{equation}
P={\displaystyle \sum_{m}^{M}}\left(T_{m\uparrow}-T_{m\downarrow}\right)\bigl/{\displaystyle \sum_{m}^{M}}\left(T_{m\uparrow}+T_{m\downarrow}\right),\label{Polarization1}\end{equation}
 where $m$ is the index and $M$ the total number of transport channels. Here,
\begin{equation}
T_{m\uparrow(\downarrow)}={\displaystyle \frac{2(k_{m}^{2}/2m^{\ast}\pm U_{0}\cos(\omega t))}{2(\hbar^{2}k_{m}^{2}/2m^{\ast}\pm U_{0}\cos(\omega t))+{\displaystyle \upsilon^{2}m^{\ast}}/\hbar^{2}},}\end{equation}
 where $k_{m}$ is the wave number of an electron in the channel $m$
at the Fermi energy. In Fig. (\ref{APolarization}), we present results
of Eq. (\ref{Polarization1}) for the maximum mechanically induced
spin-current polarization as a function of the beam width. The dashed
line in Fig. (\ref{TPolarization}) are the results of Eq. (\ref{Polarization1})
for the spin-polarization as a function of time.

The idealized model above allowed us to illustrate
the physics of piezo-spintronics. We now consider a more realistic
model, including many electron modes, subband mixing and arbitrary
forms of the potential $V(y,z)$. We numerically calculate the scattering
matrix using the recursive Green's function technique and the tight-binding
representation of the Hamiltonian (\ref{Rashba}): \[
\begin{array}{c}
H={\displaystyle \sum_{ij\sigma}\epsilon_{ij\sigma}c_{ij\sigma}^{\dagger}c_{ij\sigma}}+t\sum_{ij\sigma}\left(c_{i+1j\sigma}^{\dagger}c_{ij\sigma}+c_{ij+1\sigma}^{\dagger}c_{ij\sigma}\right)\\
\\-it_{so}\sum_{ij\sigma\sigma^{\prime}}\left(c_{i+1j\sigma}^{\dagger}c_{ij\sigma^{\prime}}(-\sin(2f)\sigma_{y})^{\sigma\sigma^{\prime}}\right.\\
\\\left.+c_{i+1j\sigma}^{\dagger}c_{ij\sigma^{\prime}}(\cos(2f)\sigma_{x})^{\sigma\sigma^{\prime}}\right)+H.c.\end{array},\]
 where $\epsilon_{ij\sigma}$ is the on-site energy that includes
$V$ and $U$, $t=\hbar^{2}/(2m^{\ast}b^{2}$) is the hopping energy
and $t_{so}=\hbar^{2}/(2l_{so}m^{\ast}b)$ is the hopping energy due
to the spin-orbit interaction, in terms of the tight-binding lattice
spacing $b$. The bold lines in Fig. \ref{TPolarization} display
our numerical results for the polarization as a function of time for
the short-range wall potential used above. We find good agreement
with the analytical results for large aspect ratios of the beam, as
expected. Deviation from the analytical results become noticeable
when $\bigtriangleup_{so}\backsim{\displaystyle \frac{\hbar\omega L}{d}}$.

\begin{figure}[t]
\centerline{\includegraphics[width=0.9\columnwidth]{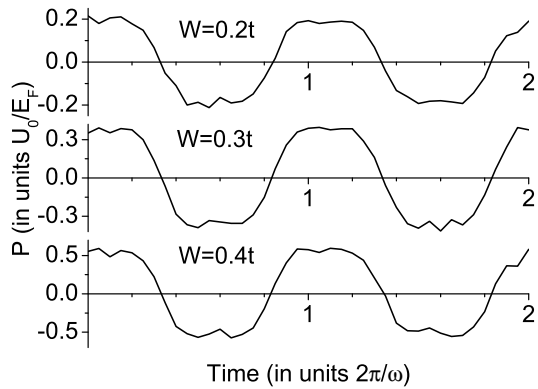}}
\vskip -0.5 cm
\caption{Spin-current polarization in units of the effective splitting $U_{0}/E_{F}$
as a function of time for the torsional oscillations of the beam in
Fig. (\ref{RodDraw}). Parameters are the same as in the figures above.}
\vskip -0.5 cm
\label{DPolarization}
\end{figure}
Finally, we model the potential $V(y,z)$ by on-site Anderson disorder
with energies distributed over a band width $W$. The beam is represented
as a $150\times9$ discrete lattice and an ensemble averaging over
$20,000$ impurity configurations is carried out. A single realization
(without averaging) behaves similar to our single defect result in
Fig. \ref{TPolarization}. Averaged results are presented in Fig.
\ref{DPolarization}. The Anderson disorder strength can be measured
by an effective 2D mean free path \citep{Ando:oct1991} as $l_{2D}=(6\lambda_{F}^{3}E_{F}^{2})/(\pi^{3}a^{2}W^{2})$.
Here we consider $l_{2D}=8L$; $3.6L$ and $2L$. When the Anderson
disorder is weak, the spin polarization is almost a harmonic function
of time. We may conclude that the piezo-spintronic effect is very
robust, persisting in a disordered system, confirming the qualitative 
behavior of the 
analytical model with one  dominating defect scatterer.

The choice of parameters above is motivated by the following estimates made
for a silicon cantilever of size $a\times d\times L=(0.05\times0.15\times1)\mu\mbox{m}$.
The resonant frequency is $\omega=c_{\mathrm{t}}\pi/L=10\,\mbox{GHz}$,
using the density of silicon $\rho=2\times10^{3}\,\mbox{kg}\,\mbox{m}^{-3}$
and the Lamé constant $\mu=100\,\mathrm{G}\mbox{Pa}$. The maximum angle
of torsion $\varphi_{0}$ can be estimated by equating the energy
dissipation during a cycle $2\pi\varphi_{0}^{2}C/(QL)$, where $Q$
is the mechanical quality factor\textbf{,} with the energy input rate
$\varphi_{0}T,$ where $T$ is the actuating torque applied at the
center of the beam. Electrostatic torques of $T\thicksim10^{-12}\div10^{-15}\mbox{Nm}$
have been already realized \citep{Fennimore:jul2003}. Taking $Q=500$,
$T=10^{-15}\mbox{Nm}$ and $C=10^{-18}\mbox{Nm}^{2}$, we find $\varphi_{0}=0.2\:\mbox{rad}$.
The corresponding spin-orbit precession length is $l_{so}=1\mu\mbox{m}$
and the spin splitting is $U_{0}={\displaystyle \frac{\hbar\omega L}{l_{so}}}\thickapprox6\times10^{-6}\mbox{eV},$
using the bulk strain-spin-orbit coupling parameter for GaAs $\gamma=2\times10^{8}\,\mbox{m}^{-1}$
\citep{Pikus:1984} which is not expected to be affected strongly by
the finite structure confinement \citep{Winkler:2003}. This leads to polarizations of the order
of $P\thicksim10^{-4}$. In the presence of an applied DC voltage,
the mechanical motion generates AC spin current. Alternatively, one
can apply AC voltage synchronized with the mechanical motion to obtain
a DC spin current. The thus created spin accumulation can be detected
by e.g. a ferromagnetic side contact \citep{Lou:may2006} or
by the optical Kerr rotation \citep{Kato:dec2004}.

In order to increase the polarization, one can use semiconductors
with lower doping (smaller Fermi energy). The width of the rod can
be tuned to capture the resonant features in Fig. \ref{APolarization}\textbf{.}
Rods with higher quality factors can have larger amplitude of oscillations
leading to higher polarizations. 

Summarizing, we propose a \textit{piezo-spintronic} effect that is
based on strain-induced coupling of the electron spin degree of freedom
and mechanical vibrations in free standing semiconductor nanobeams.
We show that time-dependent strain due to torsional mechanical oscillations
can lead to a measurable spin polarization of an applied charge current.
Mechanically generated spin-dependent potentials (mechanically-induced
Zeeman splittings) can be also used for the manipulation of an applied
spin currents. 
We propose ways to measure and increase such mechanically
generated polarization that can be used for effective spin injection
in spintronic based devices.

We thank Artem Abanov and Karel Výborný for helpful discussions. This
work was supported by ONR under
Grant No. 000140610122, by NSF under
Grant no. DMR-0547875, and by SWAN-NRI.

\bibliographystyle{apsrev} \bibliographystyle{apsrev}

\end{document}